\documentclass[a4paper,12pt]{iopart}

\usepackage{iopams, setstack}
\usepackage{setspace}
\usepackage[pdftex]{graphicx}
\usepackage{epsfig}
\usepackage{cite, hyperref}
\begin{document}
\title[Gilbert damping in Nickel thin films]{Intrinsic and non-local Gilbert damping in polycrystalline nickel studied by Ti:Sapphire laser fs spectroscopy}

\author{J Walowski$^1$, M Djordjevic Kaufmann$^1$, B Lenk$^1$, C Hamann$^2$ and J McCord$^2$, M M\"unzenberg$^1$}

\address{$^1$ Universit\"at G\"ottingen, Friedirch-Hund-Platz 1, 37077 G\"ottingen, Germany}
\address{$^2$ IFW Dresden, Helmholtzstra\ss e 20, 01069 Dresden}
\ead{walowski@ph4.physik.uni-goettingen.de}

\begin{abstract}
The use of femtosecond laser pulses generated by a Ti:Sapphire laser system allows us to gain an insight into the magnetization dynamics on time scales from sub-picosecond up to $1\,\rm{ns}$ directly in the time domain. This experimental technique is used to excite a polycrystalline nickel (Ni) film optically and probe the dynamics afterwards. Different spin wave modes (the Kittel mode, perpendicular standing spin-wave modes (PSSW) and dipolar spin-wave modes (Damon-Eshbach modes)) are identified as the Ni thickness is increased. The Kittel mode allows determination of the Gilbert damping parameter $\alpha$ extracted from the magnetization relaxation time $\tau_{\alpha}$. The non-local damping by spin currents emitted into a non-magnetic metallic layer of vanadium (V), palladium (Pd) and the rare earth dysprosium (Dy) are studied for wedge-shaped Ni films $1\,\rm{nm}-30\,\rm{nm}$. The damping parameter increases from $\alpha=0.045$ intrinsic for nickel to $\alpha>0.10$ for the heavy materials, such as Pd and Dy, for the thinnest Ni films below $10\,\rm{nm}$ thickness. Also, for the thinnest reference Ni film thickness, an increased magnetic damping below $4\,\rm{nm}$ is observed. The origin of this increase is discussed within the framework of line broadening by locally different precessional frequencies within the laser spot region. 
\end{abstract}
\section{Introduction}
The understanding of picosecond-pulsed excitation of spin packets, spin wave modes and spin currents is of importance in developing a controlled magnetic switching concept beyond the hundred picosecond timescale and to test the speed of magnetic data storage media heading to the physical limits. Over the last years profound progress has been made within that field by using femtosecond laser spectroscopy. The recent discoveries in ultrafast magnetization dynamics are heading to a new understanding \cite{Koopmans2005,Duerr2007,Chantrell2007,Chantrell2008,Djordjevic2007} 
and new all-optical switching concepts have been discovered \cite{Stanciu2007}. In addition, the all-optical method has developed into a valuable tool to study the magnetization dynamics of the magnetic precession and thereby access magnetocrystalline anisotropies and the magnetic damping \cite{Zhao2005,Talbayev2006,Rzhevsky2007,Liu2007,Muller2008} or the dynamics of magnetic modes in nanometer sized arrays of magnetic structures \cite{Lepadatu2007,Kruglyak2007} and single magnetic nanostructures \cite{Barman2006,Laraoui2007}. Naturally, one finds similarities and differences as compared to magnetic resonance techniques in frequency space (FMR) \cite{FMR_review}, optical techniques such as Brillouin light scattering (BLS) \cite{BLS_review,BLS_review_a} and time-resolved techniques, for example pulsed inductive magnetometry (PIMM) \cite{PIMM_review}. Advantages and disadvantages of the different techniques have already been compared in previous work \cite{Nibarger2006,Schneider2007,Neudecker2006}. The same concepts can be applied to the femtosecond-laser-based all-optical spectroscopy techniques. Here we discuss their abilities, highlighting some aspects and peculiarities \cite{Kampfrath2002,Djordjevic2006a,Djordjevic2006,Muller2008,Walowski2008,Lenk2008}:
\begin{itemize}
\item[i.] After excitation within the intense laser pulse, the nature of the magnetic relaxation mechanisms determine the magnetic modes observed on the larger time scale \cite{Djordjevic2007}. For a Ni wedge different modes are found as the thickness is increased: coherent precession (Kittel mode), standing spin waves (already found in \cite{Kampen2002}) and dipolar surface spin waves (Damon-Eshbach modes) appear and can be identified.

\item[ii.] Magnetic damping has been extracted by the use of fs spectroscopy experiments already in various materials, epitaxial films, as a function of the applied field strength, field orientation and laser excitation power \cite{Zhao2005,Talbayev2006,Rzhevsky2007,Liu2007,Muller2008}. Using the Kittel mode, we study the energy dissipation process caused by non-local damping by spin currents \cite{Tserkovnyak2002} in Ni by attaching a transition metal film (vanadium (V), palladium (Pd) and a rare earth film (dysprosium (Dy)) as a spin sink material and compare them to a Ni reference sample. The present advantages and disadvantages of the method are discussed.

\item[iii.] A modification of the magnetic damping is found for the thinnest magnetic layers below $4\,\rm{nm}$. The understanding of this effect is of high interest because of the increase in methods used to study magnetic damping processes in the low field region in the current literature. We present a simple model of line broadening known from FMR \cite{Heinrich1985,Celinski1991,Rantschler2003} and adapted to the all-optical geometry that pictures the effect of the increased intrinsic apparent damping observed. Therein a spread local magnetic property within the probe spot region is used to mimic the increased apparent damping for the low field region.
\end{itemize}

\begin{figure}[!h]
	\centering
		\includegraphics[width=0.75\textwidth]{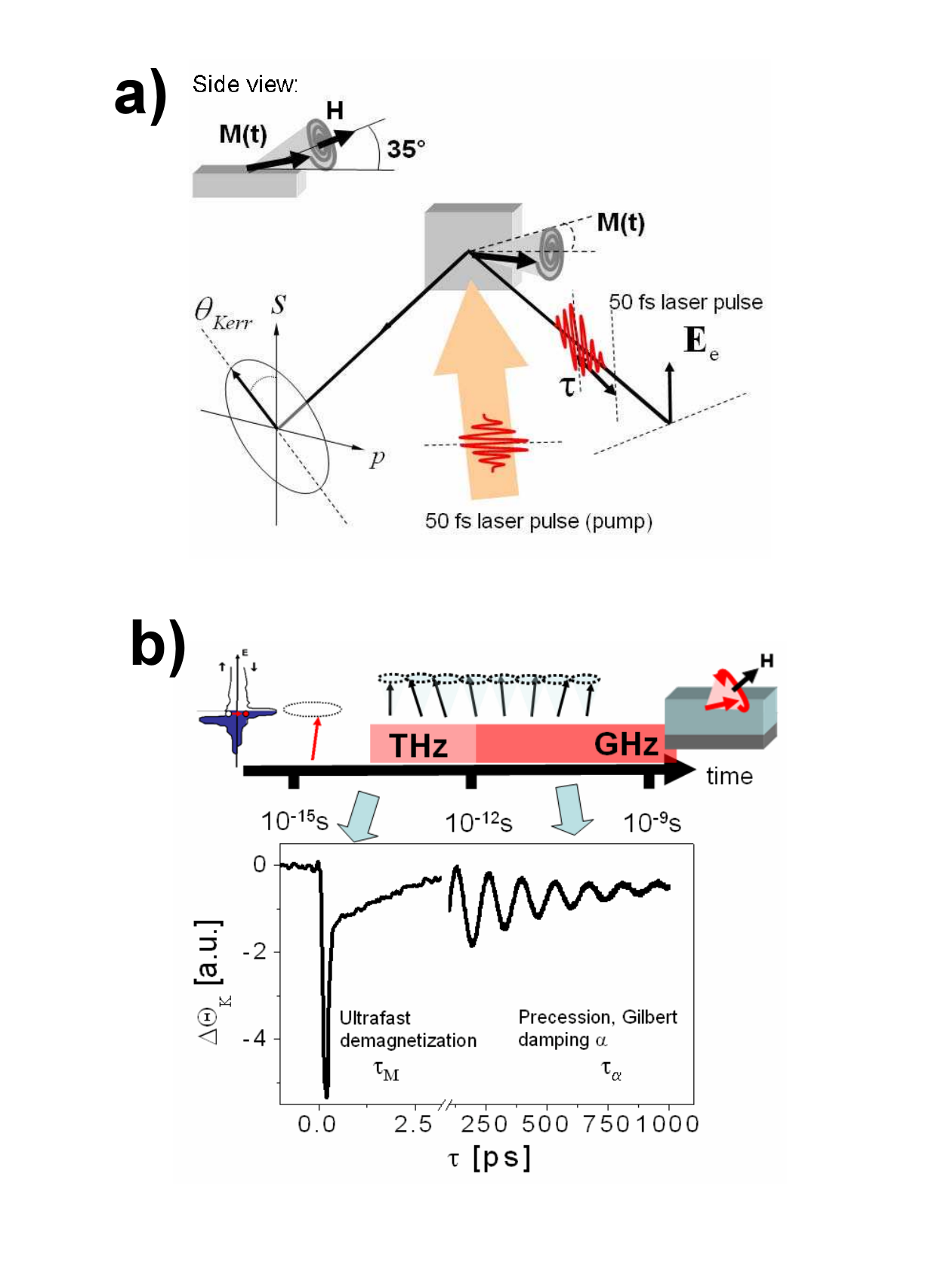}
		\caption{a) Schematics of the pump probe experiment to determine the change in Kerr rotation as a function of the delay time $\tau$. b) Experimental data on short and long time scales. On top a schematic on the processes involved is given.}
		\label{fig1}
	\end{figure}

\section{Experimental Technique}
The all-optical approach to measuring magnetization dynamics uses femtosecond laser pulses in a pump-probe geometry. In our experimental setup a Ti:Sapphire oscillator generates the fs laser pulses which are then amplified by a regenerative amplifier (RegA 9050). This system’s laser pulse characteristics are $815\,\rm{nm}$ central wave length, a repetition rate of $250\,\rm{kHz}$, a temporal length of $50-80\,\rm{fs}$ and an energy of $\sim1\,\rm{\mu J}$ per pulse. The beam is split into a strong pump beam ($95\%$ of the incoming power), which triggers the magnetization dynamics by depositing energy within the spot region, and a weaker probe pulse ($5\%$ of the incoming power) to probe the magnetization dynamics via the magneto-optical Kerr effect delayed by the time $\tau$, in the following abbreviated as time-resolved magneto-optic Kerr effect (TRMOKE). The schematic setup and sample geometry is given in figure \ref{fig1}a). The spot diameters of the pump and probe beam are $60\,\rm{\mu m}$ and $30\,\rm{\mu m}$ respectively. A double-modulation technique is applied to detect the measured signal adapted from \cite{Koopmans2000}: the probe beam is modulated with a photo-elastic modulator (PEM) at a frequency $f_1=250\,\rm{kHz}$ and the pump beam by a mechanic chopper at a frequency $f_2=800\,\rm{Hz}$. The sample is situated in a variable magnetic field ($0-150\,\rm{mT}$), which can be rotated from $0^{\circ}$ (in-plane) to $90^{\circ}$ (out of plane) direction. The degree of demagnetization can be varied by the pump fluence ($10\,\rm{mJ/cm^2}-60\,\rm{mJ/cm^2}$) to up to $20\%$ for layer thicknesses around $30\,\rm{nm}$ and up to over $80\%$ for layers thinner than $5\,\rm{nm}$. The samples studied were all grown on $\rm{Si}(100)$ substrates by e-beam evaporation in a UHV chamber at a base pressure of $\sim5\times 10^{-10}\,\rm{mbar}$. For a variation of the thickness, the layers are grown as wedges with a constant gradient on a total wedge length of $15\,\rm{mm}$.

\section{Results and discussion}
\subsection{Kittel mode, standing spin waves and Damon-Eshbach surface modes}

To give an introduction to the TRMOKE signals $\Delta \theta_{Kerr}(\tau)$ measured on the timescale from picoseconds to nanoseconds first, the ultrafast demagnetization on a characteristic time scale $\tau_{M}$ and the magnetic precessional motion damped on a time scale $\tau_{\alpha}$ is shown for a Ni film in figure \ref{fig1}b); the schematics of the processes involved on the different time scales are given on the top. The change in Kerr rotation $\Delta \theta_{Kerr}(\tau)$ shows a sudden drop at $\tau=0\,\rm{ps}$. This mirrors the demagnetization within a timescale of $\sim200\,\rm{fs}$ \cite{Beaurepaire1996,Zhang2002,Koopmans2003}. For the short time scale the dynamics are dominated by electronic relaxation processes, as described phenomenologically in the three temperature model \cite{Beaurepaire1996} or by connecting the electron-spin scattering channel with Elliot-Yafet processes, as done by Koopmans \cite{Koopmans2003} and Chantrell \cite{Chantrell2008} later. At that time scale the collective precessional motion lasting up to the nanosecond scale is initiated \cite{Kampen2002,Ju1999}: the energy deposited by the pump pulse leads to a change in the magnetic anisotropy and magnetization, and thus the total effective field. Within $\sim10\,\rm{ps}$ the total effective field has recovered to the old value and direction again. However, the magnetization, which followed the effective field, is still out of equilibrium and starts to relax by precessing around the effective field. This mechanism can be imagined as a magnetic field pulse a few picoseconds long, and is therefore sometimes called an ‘anisotropy field pulse’. The resulting anisotropy field pulse is significantly shorter than standard field pulses \cite{Jozsa2004}. This makes the TRMOKE experiment different to other magnetization dynamics experiments.

The fact that the situation is not fully described by the model can be seen in the following. Already van Kampen et al. \cite{Kampen2002} not only observed the coherent precessional mode, they also identified another mode at a higher frequency than the coherent precession mode, shifted by $\omega_{k,n} \sim 2A k^2={2A n \pi /t_{\rm{Ni}}}^2$, the standing spin wave (PSSW) mode. It originates from the confinement of the finite layer thickness, where $A$ is the exchange coupling constant and $n$ is a given order. Here we also present the finding of dipolar propagating spin waves. For all three, the frequency dependence as a function of the applied magnetic field will be discussed, a necessity for identifying them in the experiments later on.

For the coherent precession the frequency dependence is described by the Kittel equation. It is derived by expressing the effective field in the Landau-Lifshitz-Gilbert (LLG equation) as a partial derivative of the free magnetic energy \cite{Miltat2002,Farle1998}. Assuming negligible in-plane anisotropy in case of the polycrystalline nickel (Ni) film and small tilting angles of the magnetization out of the sample plane (field is applied $35^{\circ}$ out of plane figure \ref{fig1}a)), it is solved as derived in \cite{Djordjevic_thesis}: 

\begin{equation}
	\omega=\frac{\gamma}{\mu_0}\sqrt{\mu_0H_x\left(\mu_0H_x+\mu_0M_s-\frac{2K_z}{M_s}\right)},
	\label{eq1}
\end{equation}

For the standing spin waves (PSSW) a similar equation is given. For the geometry with the field applied $35^{\circ}$ out of plane (figure \ref{fig1}a) the frequencies $\omega$ and $\omega_{k,n}$ do not simply add as in the field applied in plane geometry \cite{Djordjevic_thesis}:

\begin{equation}
	\omega=\frac{\gamma}{\mu_0}\sqrt{(\mu_0H_x+\frac{2A k^2}{M_s})\left(\mu_0H_x+\mu_0M_s-\frac{2K_z}{M_s}+\frac{2A k^2}{M_s}\right)},
	\label{eq2}
\end{equation}

While the exchange energy dominates in the limit of small length scales, the magnetic dipolar interaction becomes important at larger length scales. Damon and Eshbach \cite{DamonEshbach} derived by taking into account the dipolar interactions in the limit of negligible exchange energy, the solution of the Damon-Eshbach (DE) surface waves propagating with a wave vector $q$ along the surface, decaying within the magnetic layer. The wavelengths are found to be above the $>\rm{\mu m}$ range for Ni \cite{Lenk2008}.

\begin{equation}
	\omega=\frac{\gamma}{\mu_0}\sqrt{\mu_0H_x\left(\mu_0H_x+\mu_0M_s-\frac{2K_z}{M_s}+\frac{M_S^2}{4}\left[1-\exp(-2q t_{Ni})\right]\right)},
	\label{eq3}
\end{equation}

The depth of the demagnetization by the femtosecond laser pulse is given by the optical penetration length $\lambda_{\rm{opt}}\approx15\,\rm{nm}$ ($\lambda= 800\,\rm nm$). From the nature of the excitation process in the TRMOKE experiment one can derive that for different thicknesses $t_{\rm Ni}$ it will change from an excitation of the full film for a $\sim 10\,\rm nm$ film to a thin excitation ‘layer’ only for a few $100\,\rm nm$ thick film; thus the excitation will be highly asymmetric. The model of the magnetic anisotropy field pulse fails to explain these effects since it is based on a macrospin picture.

Another way to look at the excitation mechanism has been discussed by Djordjevic et al. \cite{Djordjevic2007}. When the magnetic system is excited, on a length scale of the optical penetration depth short wavelength (high $k$ vector) spin-wave excitations appear. As time evolves, two processes appear: the modes with high frequency owning a fast oscillation in space are damped very fast by giving part of the deposited energy to the lattice. In addition, through multiple magnon interaction lower $k$-vector states are populated, resulting in the highest occupation of the lowest energy modes at the end (e.g. the PSSW and DE modes here). As the Ni thickness is increased, the excitation profile becomes increasingly asymmetric, favoring inhomogeneous magnetic excitations, as the PSSW mode. The DE modes, due to their nature based on a dipolar interaction, are expected to be found only for higher thicknesses.

\begin{figure}[!h]
	\centering
		\includegraphics[width=1.00\textwidth]{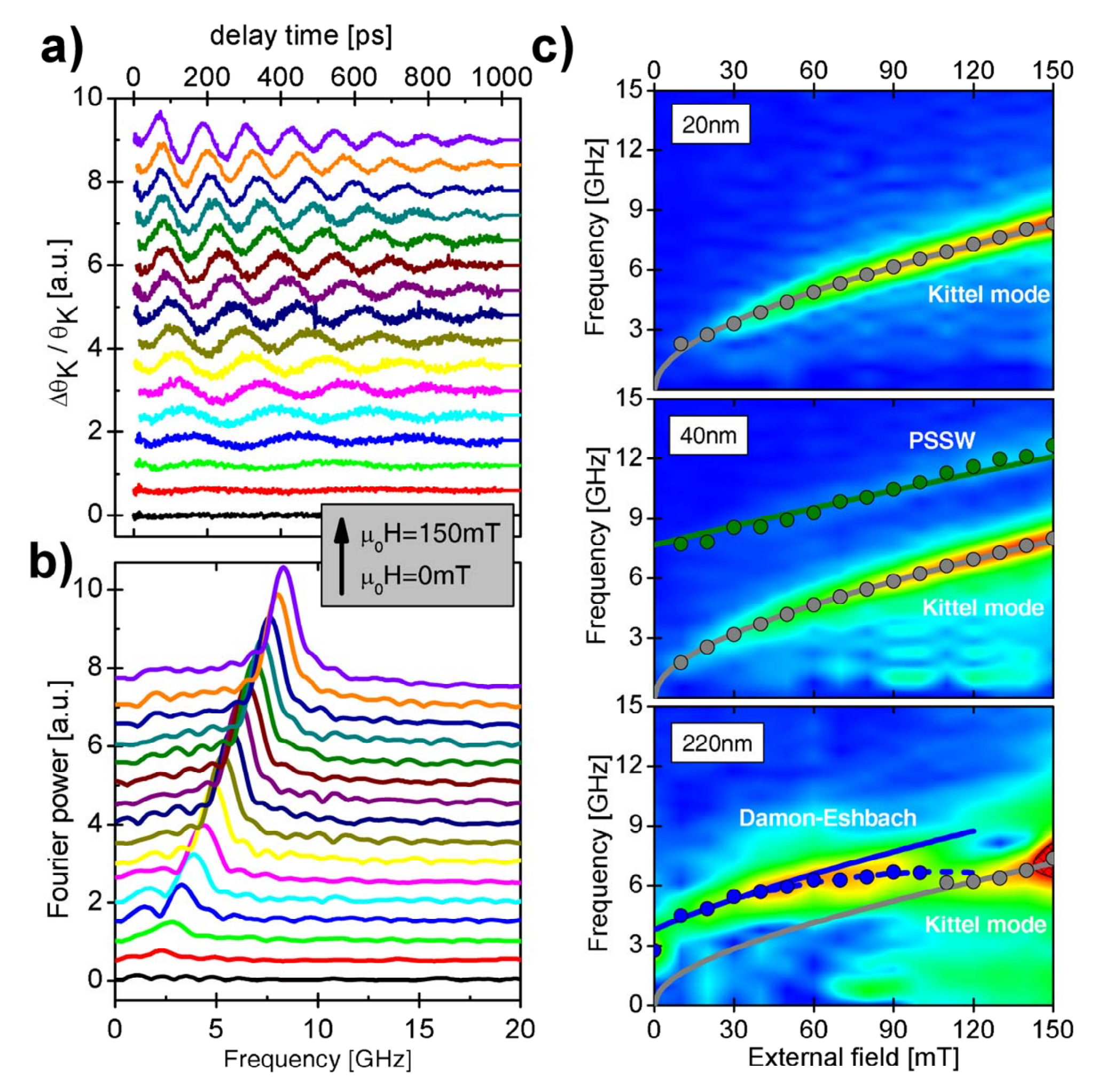}
	\caption{Change in Kerr rotation after excitation on the long time scale for $\rm{Cu}\,2\rm{nm}/$ $\rm{Ni}\,{t_{\rm Ni}}\,\rm{nm}/\rm{Si}(100)$ with $t_{\rm Ni}=20$ and b) their Fourier transform for different applied fields $0-150\,\rm mT$, ($35^{\circ}$ out of plane (blue)). In c) the Fourier power spectra as color maps for three Ni thicknesses $t_{\rm Ni}=20$, $40$ and $220\,\rm nm$ are given. The data overlaid is determined form the peak positions. The straight lines are the analysis of the different modes and are identified in the graph (Kittel model), perpendicular standing spin wave (PSSW) and dipolar surface spin wave (Damon Eshbach mode).}
	\label{fig2}
\end{figure}

The identification of the mode is important in determining a value for the magnetic damping $\alpha$. Figure \ref{fig2} pictures the identification of the different modes and their appearance for different Ni thicknesses. The data are handled as follows: for a $t_{\rm Ni}=20\,\rm nm$ film on Si(100), covered with a $2\,\rm nm$ Cu protection layer, in a) the original data after background subtraction and in b) its corresponding Fourier transform, shown for increasing applied magnetic field. The evolution of the mode frequency and its amplitude increase can be followed. An exponentially decaying incoherent background is subtracted from the data. This has to be done very carefully, to avoid a step-like background which will be evident after Fourier transform as a sum of odd higher harmonics. The frequency resolution is limited by the scan range of $1\,\rm ns$ corresponding to $\Delta\omega/2\pi=1\,\rm  GHz$. However, since the oscillation is damped within the scan range, the datasets have been extended before Fourier transform to increase their grid points. A color map of the power spectrum is shown in figure \ref{fig2}c), where the peak positions are marked by the data points overlaid. For the $20\,\rm nm$ thick film with $t_{\rm Ni}=20\,\mathrm{nm}\sim\lambda_{\rm opt}$ only a single mode is observed. The mode is analyzed by \ref{eq1} indicating the Kittel mode being present (data points and line in figure \ref{fig2}c), top) using $K_z=3.03 \cdot 10^{4}\,\rm{J/m^3}$. With increasing nickel thickness $t_{Ni}=40\,\mathrm{nm}>\lambda_{\rm opt}$, the perpendicular standing spin waves (PSSW) of first order are additionally excited and start to appear in the spectra (figure \ref{fig2}c), middle). An exchange constant $A=9.5\cdot 10^{12}\,\rm{J/m}$ is extracted. In the limit of $t_{\rm Ni}=220\,\rm nm \gg\lambda_{\rm opt}$ (figure \ref{fig1}c)) the excitation involves the surface only. Hence, modes with comparable amplitude profile, e.g. with their amplitude decaying into the Ni layer, are preferred. Consequently DE surface waves are identified as described by \ref{eq3} and dominate the spectra up to critical fields as high as $\mu_0H_{\rm crit}=100\,\rm mT$. For $t_{\rm Ni}=220\,\rm nm$ the wave factor is $k=2\,\mu\rm{m}$ (data points and line in figure \ref{fig2}c), bottom). For larger fields than $100\,\rm mT$ the DE mode frequency branch merges into the Kittel mode \cite{Lenk2008}.

To resume the previous findings for the first subsection, we have shown that in fact the DE modes, though they are propagating spin-wave modes, can be identified in the spectra and play a very important role for Ni thicknesses above $t_{\rm Ni}=80\,\rm nm$. They appear for thicknesses much thinner than the wavelength of the propagating mode. Perpendicular standing spin waves (PSSW) give an important contribution to the spectra for Ni thicknesses above $t_{\rm Ni}=20\,\rm nm$. For thicknesses below $t_{\rm Ni}=20\,\rm nm$ we observe the homogeneously precessing Kittel mode only. This thickness range should be used to determine the magnetic damping in TRMOKE experiments.

\subsection{Data analysis: determination of the magnetic damping}

\begin{figure}[!h]
	\centering
		\includegraphics[width=1.00\textwidth]{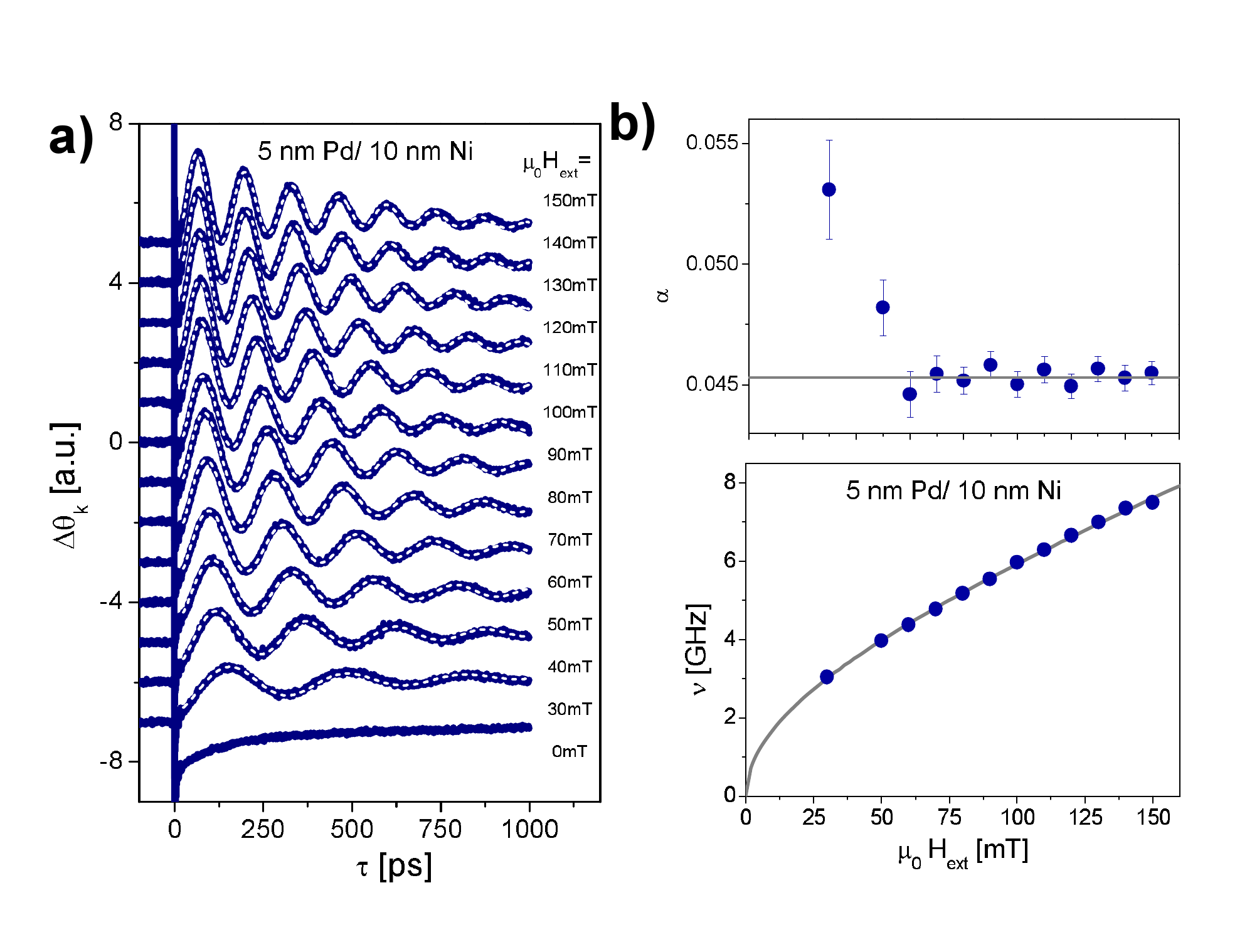}
	\caption{a) Kerr rotation spectra for a $\rm{Cu}\,2\rm{nm}/$ $\rm{Ni}\,10\,\rm{nm}/$ $\rm{Pd}\,5\,\rm{nm}/\rm{Si}(100)$ layer, measured for fields applied from $0-150\,\rm mT$ ($35^{\circ}$ out of plane, (blue)) and the fitted functions (white, dashed). b) The magnetic damping $\alpha$ and precession frequencies extracted from the fits to the measured spectra. The line is given by the Kittel mode (gray).}
	\label{fig3}
\end{figure}

For the experiments carried out in the following with $t_{\rm Ni}<25\,\rm nm$ the observed dynamics can be ascribed to the coherent precession of the magnetization (Kittel mode). The analysis procedure is illustrated in the following using the data given in figure \ref{fig3}a). A Pd layer is attached to a Ni film with the thicknesses ($\rm{Ni}\,10\,\rm{nm}/$ $\rm{Pd}\,5\,\rm{nm}/$ $\rm{Si}(100)$) to study the non-local damping by spin currents absorbed by the Pd. The different spectra with varying the magnetic field strength from $0\,\rm{mT}-150\,\rm{mT}$ are plotted from bottom to top (with the magnetic field tilted $35^{\circ}$ out of the sample plane).

The data can be analyzed using the harmonic function with an exponential decay within $\tau_{\alpha}$:

\begin{equation}
	\Delta\theta_k\sim\exp\left(-\frac{\tau}{\tau_{\alpha}}\right)\cdot\sin(2\pi(\tau-\tau_0)\nu)+B(\tau),
	\label{eq4}
\end{equation}

The precession frequency $\nu=\omega/2\pi$ and the exponential decay time $\tau_{\alpha}$ of the precession amplitude is extracted, where the function $B(\tau)$ stands for the background arising from the uncorrelated magnetic and phonon excitations. To determine the Gilbert damping parameter $\alpha$ as given in the ansatz by Gilbert, the exponential decay time $\tau_{\alpha}$ has to be related with $\alpha$. The LLG equation is solved under the same preconditions as for equation \ref{eq1} using an exponential decay of the harmonic precession within $\tau_{\alpha}$ from \ref{eq4}. Then the damping parameter $\alpha$ and can be expressed by the following equation \cite{Djordjevic_thesis}:

\begin{equation}
	\alpha=\frac{1}{\tau_{\alpha}\gamma\left(H_x-\frac{K_z}{\mu_0M_s}+\frac{M_s}{2}\right)}.
	\label{eq5}
\end{equation}

It is evident from \ref{eq5} that in order to determine the Gilbert damping $\alpha$ from the decay of the Kittel mode $\tau_{\alpha}$, the variables $\gamma$, $M_s$ and $K_z$ have to be inserted, and therefore  $K_z$ has to be determined beforehand.

In figure \ref{fig3}a), the background $B(\tau)$ is already subtracted. The fits using \ref{eq4} are plotted with the dashed lines on top of the measured spectra. The results are presented in b). The frequencies range from $3\,\rm{GHz}$ for $30\,\rm{mT}$ to $7.5\,\rm{GHz}$ for the $150\,\rm mT$ applied magnetic field. They increase linearly with the strength of the applied magnetic field for high field values. The extrapolated intersection with the ordinate is related to the square root of the dipolar and anisotropy field. Using the Kittel equation (\ref{eq1}), one determines the out-of-plane anisotropy constant $K_z$ of $K_z=6.8\cdot10^{4}\,\rm{J/m^3}$. The calculated magnetic damping $\alpha$ as a function of the applied field is given in the graph below: this is mostly constant but increases below $60\,\rm mT$. Within the ansatz given by Gilbert, the damping constant $\alpha$ is assumed to be field-independent. We find that this is fulfilled for most of the values: the average value of $\alpha=0.0453(4)$, consistent with earlier findings by Bhagat and Lubitz from FMR experiments \cite{Bhagat1974}, is indicated by the line in the plot. The $\alpha$ given in the following will always be averaged over a field region where the damping is Gilbert-like. A deviation from this value occurs for the small external field strengths. It originates for two reasons: fitting \ref{eq5} with a few periods only does not determine a reliable value of the exponential precession decay time $\tau_{\alpha}$ and leads to a larger error. Second, magnetic inhomogeneities mapping a spread in anisotropy energies within the probe spot region can also be a source, and this becomes generally more important for even thicker films below $4\,\rm{nm}$ \cite{Celinski1991}. This will be discussed in more detail in the last section of the manuscript.

\subsection{Intrinsic damping: nickel wedge}

\begin{figure}[!h]
	\centering
		\includegraphics[width=1.00\textwidth]{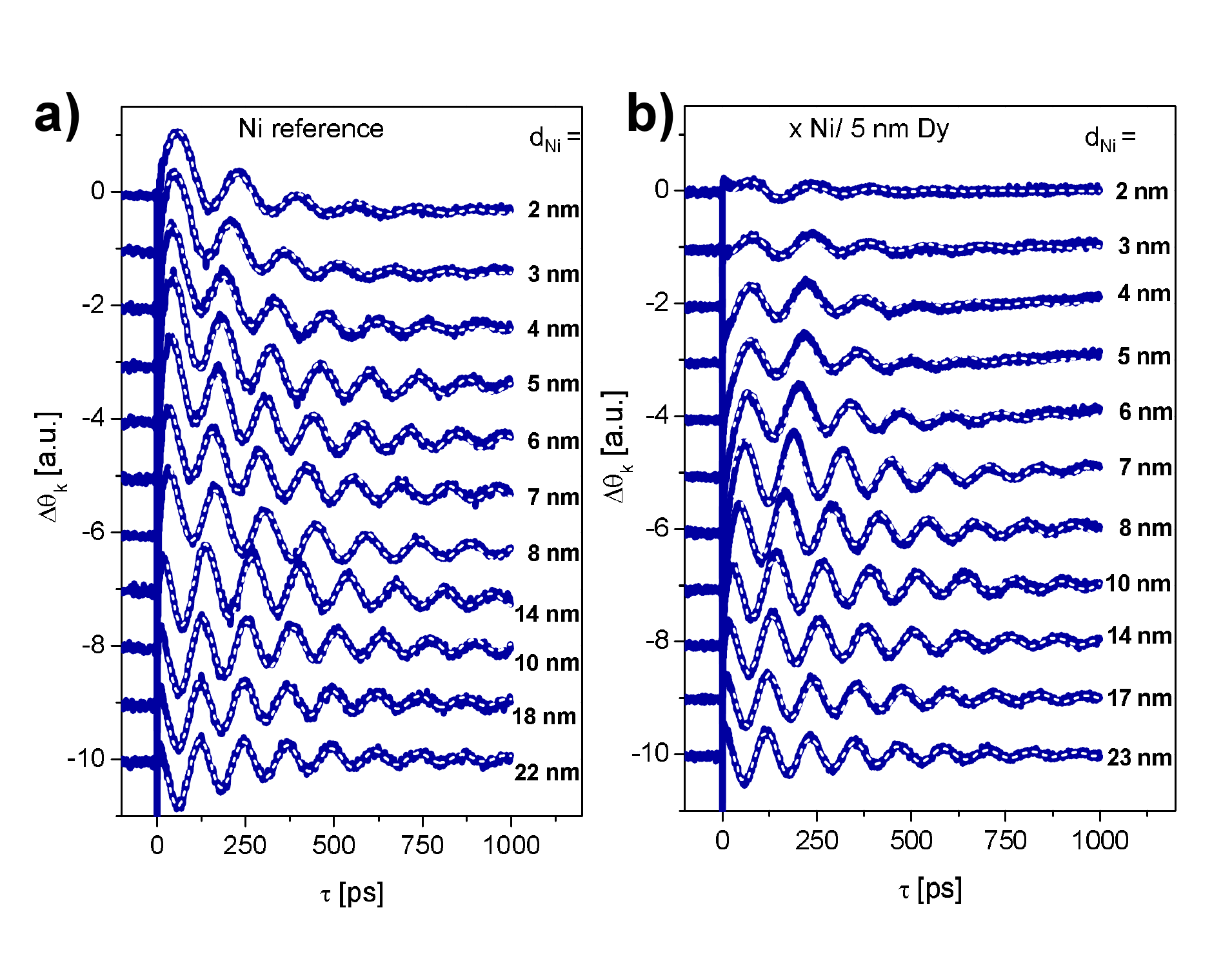}
	\caption{a) Kerr rotation spectra for nickel layers from $t_{\rm Ni}=2\,\rm{nm}-22\,\rm{nm}$, measured on the nickel wedge $t_{\rm Ni}=\,\rm{nm}\,\rm{Ni}/\rm{Si}(100)$ and opposed in b) by a nickel wedge $\rm{Al}\,2\rm{nm}/$ $\rm{Dy}\,5\rm{nm}/$ $t_{Ni}=\,\rm{nm}\,\rm{Ni}/\rm{Si}(100)$ with a $5\,\rm nm$ Dy spin-sink layer.}
	\label{fig4}
\end{figure}

For our experiments Ni was chosen instead of Fe or Py as a ferromagnetic layer. The latter would be preferable because of their lower intrinsic damping $\alpha_{\rm int}$, which make the films more sensitive for detecting the non-local contribution to the damping. The reason for using Ni for our experiments is the larger signal excited in the TRMOKE experiments. The magnetic damping $\alpha_{\rm int}$ is used as a reference later on. The different spectra with varying the Ni thickness $t_{\rm Ni}$ $\mathrm{Ni}\,x\,\rm{nm}/\rm{Si}(100)$ from $2\,\rm{nm}\leq x\le22\,\rm{nm}$ are plotted from bottom to top (with the constant magnetic field $150\,\rm{mT}$ and tilted $30^{\circ}$ out of plane) in figure \ref{fig4}a). The measurements were performed immediately after the sample preparation, in order to prevent oxidation on the nickel surface caused by the lack of a protection layer (omitted on purpose). The spectra show similar precession frequency and initial excitation amplitude. However, the layers with $t_{\rm Ni}<10\,\rm nm$ show a frequency shift visually recognized in the TRMOKE data. Furthermore, the precession amplitude decreases faster for the thinner layers. Figure \ref{fig5} shows the frequencies and the damping parameter extracted from the measured data in the intrinsic case for the nickel wedge sample (black squares). While the precession frequency given for $150\,\rm mT$  is almost constant above $8\,\rm nm$ Ni thickness, it starts to drop by about $25\%$ for the thinnest layer. The magnetic damping $\alpha$ (black squares) is found to increase to up to $\alpha=0.1$, an indication that in addition to the intrinsic there are also extrinsic processes contributing. It has to be noted that the change in $\alpha$ is not correlated with the decrease of the precession frequency. The magnetic damping $\alpha$ is found to increase below a thickness of $4\,\rm{nm}$, while the frequency decrease is observed below a thickness of $10\,\rm nm$. A priori $\gamma$, $M_s$ and $K_z$ can be involved in the observed frequency shift, but they can not be disentangled within a fit of our field-dependent experiments. However, from our magnetic characterization no evidence of a change of $\gamma$ and $M_s$ is found. A saturation magnetization $\mu_0M_s=0.659\,\rm T$ and $g$-factor of 2.21 for Ni are used throughout the manuscript\footnote{An altered g-factor by interface intermixing can not decrease its value below $\sim$2. Also, there is no evidence for a reduced $M_s$ for lower thicknesses found in the Kerr rotation versus Ni thickness data. More expected is a change in the magnetic anisotropy $K_z$. For the calculation of $\alpha$ later on, the in both cases (assuming a variation of $K_z$ or an altered $\gamma$) the differences are negligible.} and $K_z$ is determined as a function of the Ni thickness, which shows a $1/t_{\rm Ni}$ behavior, as expected for a magnetic interface anisotropy term \cite{Walowski2007}.

The knowledge of the intrinsic Gilbert damping $\alpha_{\rm int}$ of the Ni film of a constant value for up to $3\,\rm nm$ thickness allows us to make a comparative study of the non-local damping $\alpha'$, introduced by an adjacent layer of vanadium (V) and palladium (Pd) as representatives for transition metals, and dysprosium (Dy) as a representative of the rare earths. Both damping contributions due to intrinsic $\alpha_{\rm int}$ and non-local spin current damping $\alpha'$ are superimposed by:

\begin{equation}
	\alpha=\alpha_{int}+\alpha'.
	\label{eq6}
\end{equation}

They have to be disentangled by a study of the thickness dependence and compared to the theory of spin-current pumping, plus a careful comparison to the intrinsic value $\alpha_{\rm int}$ has to be made.

\subsection{Non-local spin current damping: theory}

Dynamic spin currents excited by a precessing moment in an adjacent nonmagnetic layer ($\rm NM$) are the consequence of the fact that static spin polarization at the interface follows a dynamic movement of a collective magnetic excitation. The effect has already been proposed in the seventies \cite{Silsbee1979,Janossy1976} and later calculated within a spin reservoir model with the spins pumped through the interfaces of the material by Tserkovnyak \cite{Tserkovnyak2002,Tserkovnyak2004}. For each precession, pumping of the spin current results in a corresponding loss in magnetization, and thus in a loss of angular momentum. The spin information is lost and the backward diffusion damps the precession of the magnetic moment. In addition to the first experiments using ferromagnetic resonance (FMR) \cite{Mizukami2001,Mizukami2001a,Mizukami2002,Foros2005,Urban2001,Urban2002,Lenz2004} it has been observed in time- resolved experiments using magnetic field pulses for excitation \cite{Lagae2005,Woltersdorf2005}. In fact the non-local spin current damping is very closely related to the damping by spin-flip scattering described within the s-d current model \cite{Mitchell1957,Heinrich1967} that uses the approximation of strongly localized d-states and delocalized s-states \cite{Stearns1977}.

A review describes the underlying circuit theory and dynamics of the spin currents at interfaces in detail \cite{Tserkovnyak2005}. The outcome of the theoretical understanding is that the additional Gilbert damping is proportional to the angular momentum $A_{r,l}$ transmitted through the interface. Since each interface owns a characteristic reflection and transmission, the size of $A_{r,l}$ depends on the matching of the Fermi surfaces. The absolute value is given by the total balance between transmitted angular momentum and the back flow. For the non-local damping $\alpha'$ one finds:

\begin{equation}
	\alpha'=\frac{\gamma\hbar G^{\uparrow\downarrow}}{4\pi M_s t_{\rm FM}}\frac{1}{1+\sqrt{\frac{\tau_{\rm sf}}{\tau_{\rm el}}}\tanh\left(\frac{t_{\rm NM}}{\lambda_{\rm sd}}\right)^{-1}}.
	\label{eq7}
\end{equation}

The $\tanh$ function stems thereby from the diffusion profile of the spin currents determined by the spin diffusion length $\lambda_{\rm sd}$ within the non-magnetic material with thickness $t_{\rm NM}$. Also, one finds from the analysis the ratio of the electron scattering rate $\tau_{\rm el}$ versus the spin flip rate $\tau_{\rm sf}$. The total amount of spin current through the interfaces is determined by the interface spin mixing conductance $G^{\uparrow\downarrow}$. It is related to the magnetic volume. It is therefore that scales with the thickness of the magnetic layer $t_{\rm FM}$. The effective gyromagnetic ratio altered by the spin-current implies that in addition to an increased damping a small frequency shift will be observed. The non-local Gilbert damping becomes important when it exceeds the intrinsic damping $\alpha_{\rm int}$. 

\subsection{Non-local damping: vanadium, palladium and dysprosium}

Differing from other techniques, TRMOKE experiments require optical access for excitation and detection, setting some restrictions to the layer stack assembly that can be investigated with this method: a thick metallic layer on top of the magnetic layer is not practical. Placing the damping layer below the magnetic layer is also unfavorable: by increasing the spin sink thickness the roughness of the metal film will increase with the metal’s layer thickness and introduce a different defects density, altering $\alpha_{\rm int}$. In the following the nickel thickness will be varied and the spin sink thickness will be kept fixed at $5\,\rm nm$. To warrant that the nickel films magnetic properties are always comparable to the reference experiment ($K_z$, $\alpha_{\rm int}$), they are always grown first on the Si(100). For the Pd case the damping layer is below the Ni layer. Here the excitation mechanism did not work and the oscillations were too weak in amplitude to analyze the damping $\alpha$, probably due to the high reflectivity of Pd.

The results are presented in figure \ref{fig4}b) for the nickel wedge sample $\mathrm{Ni}\,x\,\rm{nm}/\rm{Si}(100)$ with a $5\,\rm{nm}$ dysprosium (Dy) as a spin sink layer, covered by an aluminum protection layer, as opposed to the nickel wedge sample data without this in a). The nickel layer thickness is varied from $2\,\mathrm{nm}\leq x\le22\,\rm{nm}$. All spectra were measured in an external magnetic field set to $150\,\rm mT$ and tilted $30^{\circ}$ out of plane. For the thinnest Ni thickness, the amplitude of the precession is found to be smaller due to the absorption of the Dy layer on the top. While the precession is equally damped for the Ni thicknesses ranging from $7$ to $23\,\rm nm$, an increased damping is found for smaller thicknesses below this. The difference in damping of the oscillations is most evident for $t_{\rm Ni}=4$ and $5\,\rm nm$.

\begin{figure}[h!]
	\centering
		\includegraphics[width=0.70\textwidth]{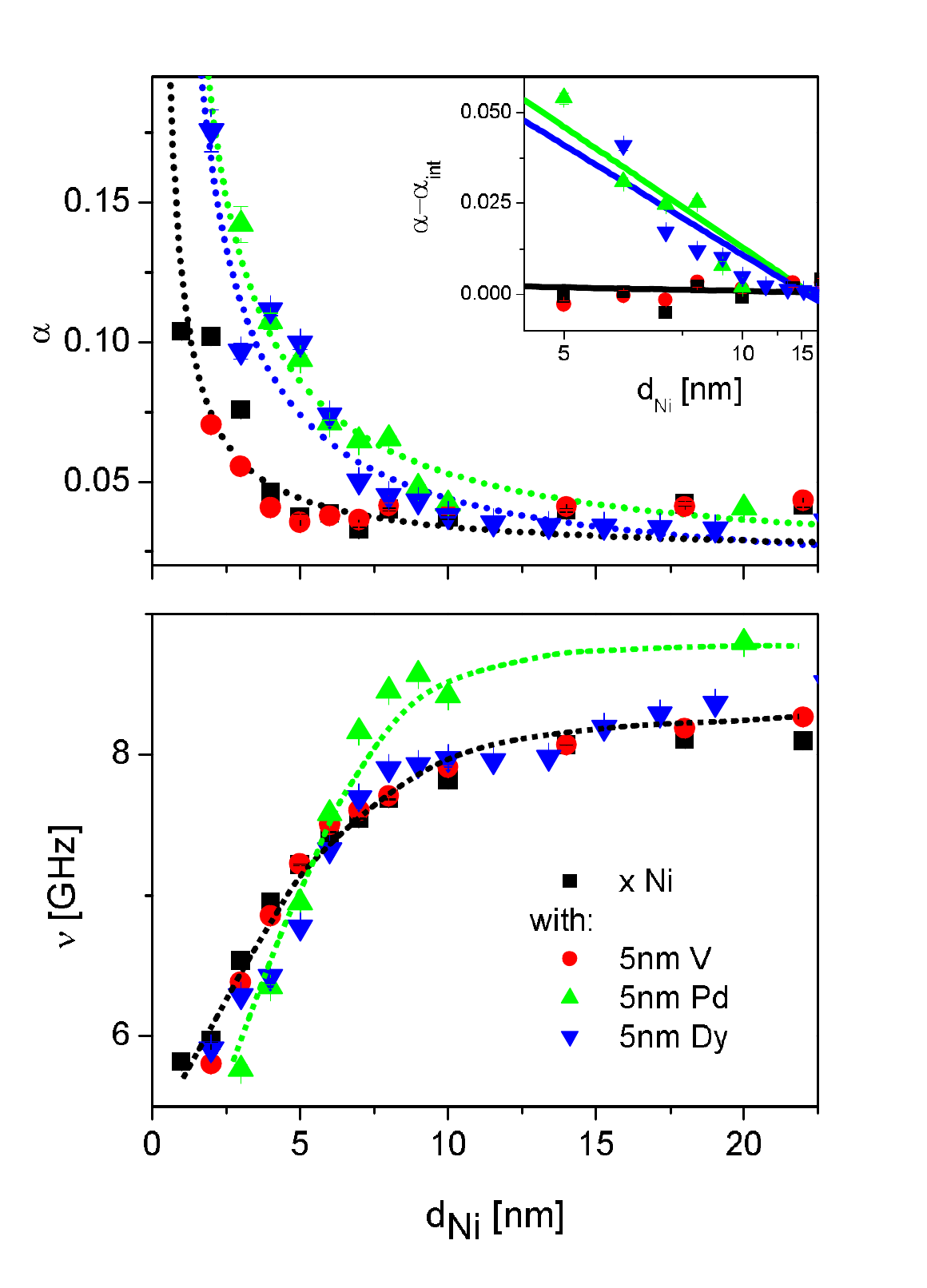}
	\caption{Gilbert damping parameters $\alpha$ and frequency $\nu$ as a function of the nickel layer thickness for the intrinsic case and for different damping materials of $5\,\rm{nm}$ V, Pd, and Dy adjacent to the ferromagnet. $\alpha$ is extracted from experiments over a large field region. The fits are made using equation 5 and equation 7. In the inset the data is shown on a reciprocal scale. Below, the frequency is given ($150\,\rm mT$). The lines are guides for the eye.}
	\label{fig5}
\end{figure}

The result of the analysis as described before is summarized in figure \ref{fig5}. In this graph the data are shown for the samples with the $5\,\rm nm$ V, Pd, Dy spin-sink layer and the Ni reference. While for the Ni reference, and Ni with adjacent V and Dy layer, the frequency dependence is almost equal, indicating similar magnetic properties for the different wedge-like shaped samples, the frequency for Pd is found to be somewhat higher and starts to drop faster than for the others. The most probable explanation is that this difference is due to a slightly different anisotropy for the Ni grown on top of Pd in this case. Nevertheless, the magnetic damping found for larger thicknesses $t_{\rm Ni}$ is comparable with the Ni reference. In the upper graph of figure \ref{fig5} the Gilbert damping as a function of the Ni layer thickness is shown. While for the Pd and Dy as a spin sink material a additional increase below $10\,\rm nm$ contributing to the damping can be identified, for V no additional damping contribution is found.

For the adjacent V layer, since it is a transition metal with a low spin orbit-scattering (light material with low atomic number Z), with a low spin-flip scattering rate and thus a spin diffusion length larger than the thickness $t_{\rm NM}$ ($d \ll \lambda_{\rm sd}$), no additional damping will occur. For Pd and Dy the situation is different: whereas the heavier Pd belongs to the transition metals with a strong orbit-scattering (heavy material with high atomic number Z), Dy belongs to the rare earth materials. It owns a localized 4f magnetic moment: therefore, both own a high spin-flip scattering rate and we expect the latter two to be in the region where ($t \gg \lambda_{\rm sd}$). In their cases the thickness of $5\,\rm nm$ of the spin-sink layer is chosen to be larger than the spin diffusion length ($t_{\rm NM} \gg \lambda_{\rm sd}$). In this limit the spin current emitted from the magnetic layer through the interface is totally absorbed within the non-magnetic layer. One can simplify \ref{eq6} to: 

\begin{equation}
	\alpha'(\infty)= \frac{\gamma\hbar G^{\uparrow\downarrow}}  {4\pi M_s} t^{-1}_{\rm FM}.
	\label{eq8}
\end{equation} 

This is called the limit of a ‘perfect spin sink’. The additional non-local spin current damping is expected to behave inversely proportional with the nickel layer thickness $\sim t^{-1}_{\rm FM}$. The inset gives the analysis and the data point on a reciprocal scale. The slope shows a linear increase for thinner nickel layers, as expected for an inverse proportionality for both the Pd and the Dy. Since the value for the intrinsic damping of the nickel film increases below $4\,\rm nm$ this contribution has to be subtracted to reveal the spin-current contribution. The value for $\alpha'$ is then found to be $0.07$ for the $2\,\rm{nm}\,$ $\rm Ni/5\,\rm{nm}\,Pd$ film, which is in the order found by Mizukami by FMR for sputtered Permalloy films with a Pd spin sink ($\alpha'=0.04$ for $2\,\rm{nm}\,Py/$ $5\,\rm{nm}\,Pd$) \cite{Mizukami2001a,Mizukami2002}. A further analysis of the thickness dependence of $\alpha$ yields values for the prefactor in \ref{eq7} for Pd ($0.33(3)\,\rm nm$) and Dy ($0.32(3)\,\rm nm$) with the fit given in the graph. From that value the real part of the interface spin mixing conductance in \ref{eq7} can be calculated. It is found to be $G^{\uparrow\downarrow}=4.5(5) \cdot 10^{15}\,\rm \Omega cm^1$ for the Ni/Pd and Ni/Dy interface. The increase of the intrinsic damping $\alpha_{\rm int}$ has been analyzed using an inverse thickness dependence (prefactor of $0.1\,\rm nm$). While it describes the data in the lower thickness range, it can be seen that it does not describe the thickness dependence for the thicker range and thus, probably the increase does not originate from an interface effect.

\subsection{Increased damping caused by anisotropy fluctuations: consequences for the all-optical approach}

\begin{figure}[h!]
	\centering
		\includegraphics[width=0.60\textwidth]{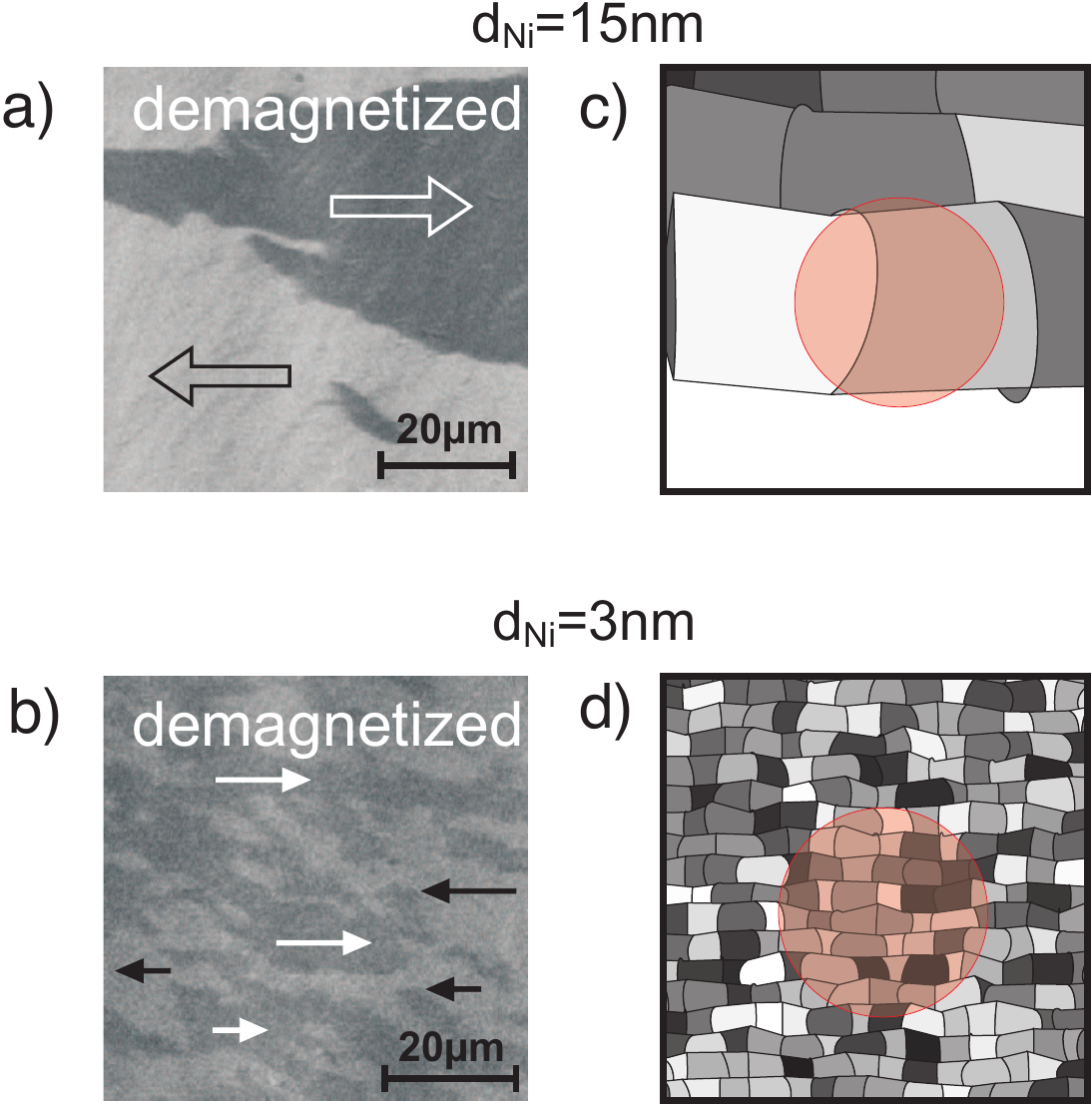}
	\caption{a) and b) Kerr microscopy images for the demagnetized state for $15\,\rm nm$ and $3\,\rm nm$. c) and d) corresponding model representing the areas with slightly varying anisotropy}
	\label{fig6}
\end{figure}

In this last part we want to focus on the deviation from the intrinsic damping $\alpha_{\rm int}$ for the thin nickel layers itself $(t_{\rm Ni}<4\,\rm nm$). In the low field range ($10-50\,\rm mT$) small magnetization inhomogeneities can build up even when the magnetization appears to be still saturated from the hysteresis curve (the saturation fields are a few mT). For these thin layers the magnetization does not align parallel in an externally applied field any more, but forms ripples. The influence of the ripples on the damping is discussed in reference \cite{Rantschler2003}. In the following we adopt this ansatz to the experimental situation of the TRMOKE experiment. We deduce a length scale on which the magnetization reversal appears for two different Ni thicknesses and relate it to the diameter of our probe spot. Lateral magnetic inhomogeneities were studied using Kerr microscopy at different applied magnetic fields \cite{Walowski2007}. Magnetization reversal takes place at low fields of a -0.5 to $2\,\rm mT$. The resolution of the Kerr microscopy for this thin layer thickness does not allow us to see the extent of the ripple effect in the external field where the increase of $\alpha$ and its strong field dependence is observed. However, the domains in the demagnetized state also mirror local inhomogeneities. For our $\mathrm{Ni}\,x\,\rm{nm}/\rm{Si}(100)$ sample this is shown in figure \ref{fig6}a) and b). The domains imaged using Kerr microscopy are shown for a $3\,\rm nm$ and a $15\,\rm nm$ nickel layer in the demagnetized state. The domains of the $15\,\rm nm$ layer are larger than the probe spot diameter of $30\,\rm{\mu m}$, whereas the domains of the $3\,\rm nm$ layer are much smaller.

From that observation, the model of local anisotropy fluctuations known from FMR \cite{Heinrich1985,Celinski1991} is schematically depicted in figure \ref{fig6}c) and d). A similar idea was also given by McMichael \cite{McMichael2007} and studied using micromagnetic simulations. While for the thick film the laser spot probes a region of almost homogeneous magnetization state, for the thin layer case the spot averages over many different regions with slightly different magnetic properties and their magnetization slightly tilted from the main direction averaging over it. The TRMOKE signal determined mirrors an average over the probed region. It shows an increased apparent damping $\alpha$ and a smaller $\tau_{\alpha}$ resulting from the line broadening and different phase in frequency space. While for the thick layer the typical scale of the magnetic inhomogeneity is as large as the probe laser spot given and only 1-2 regions are averaged, for the thinner film of  $d_{\rm Ni}=3\,\rm nm$ many regions are averaged within a laser spot, as can be seen in \ref{fig6}b) and d). Because the magnetic inhomogeneity mapping local varying anisotropies becomes more important for smaller fields, it also explains the strong field dependence of $\alpha$ observed within that region.

\begin{figure}[!h]
	\centering
		\includegraphics[width=0.55\textwidth]{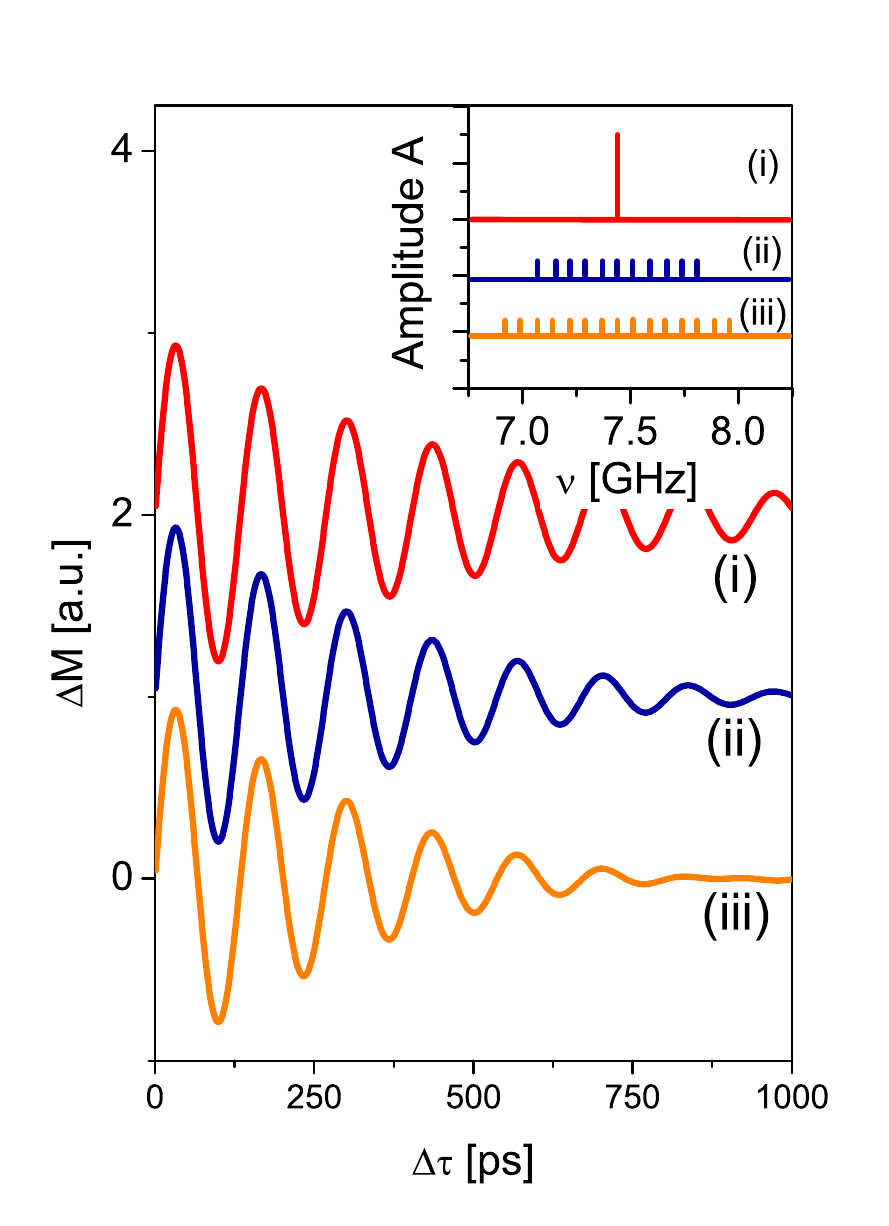}
	\caption{a) Datasets generated by superposing the spectra with the frequency spread according to the inset: (i) is calculated from the values extracted from the experimental data for the $10\,\rm nm$ Ni layer, (ii) by a superposition of spectra with up to $5\%$ and (iii) is calculated by a superposition of spectra owing $7\%$ variation from the central frequency at maximum. The average precession amplitude declines faster if a higher spread of frequencies (i.e. different anisotropies) are involved.}
	\label{fig7}
\end{figure}

Figure \ref{fig7} shows data calculated based on the model, in which the upper curve (i) is calculated from the values extracted from the experimental data for the $10\,\rm nm$ nickel layer, curve (ii) is calculated by a superposition of spectra with up to $5\%$ deviation from the central frequency at maximum and curve (iii) is calculated by a superposition of spectra of $7\%$ deviation from the central frequency at maximum to mimic the line broadening. The corresponding amplitudes of the superposed spectra related to different $K_z$ values is plotted in the inset of the graph to the given frequencies. The apparent damping is increased by 0.01 (for $5\%$) and reaches the value given in figure \ref{fig3}b) for the $10\,\rm nm$ film determined for the lowest field values of $30\,\rm mT$. These effects generally become more important for thinner films, since the anisotropy fluctuations arising from thickness variations are larger, as shown by the Kerr images varying on a smaller length scale. These fluctuations can be vice versa determined by the analysis.

\section{Conclusion}
To conclude, we have shown that all-optical pump-probe experiments are a powerful tool to explore magnetization dynamics. Although the optical access to the magnetic layer allows an access to the surface only, magnetization dynamics can be explored directly in the time domain, resolving different types of spin-wave modes (Kittel mode, perpendicular standing spin waves and Damon-Eshbach dipolar surface waves). This is in contrast to FMR experiments, where the measured data is a response of the whole sample. The obtained data can be similar to the field-pulsed magnetic excitations and the Gilbert damping parameter $\alpha$, needed for the analysis of magnetization dynamics and the understanding of microscopic energy dissipation, can be determined from these experiments. We have evaluated the contributions of non-local spin current damping for V, Pd and Dy. Yet there are limits, as shown for the nickel layer thicknesses below $4\,\rm nm$, where the magnetic damping increases and may overlay other contributions. This increase is attributed to an inhomogeneous line broadening arising from a strong sensitivity to local anisotropy variations. Further experiments will show whether the reduction of the probe spot diameter will improve the results by sensing smaller areas and thus reducing the line broadening.

\ack
The financial support by the Deutsche Forschungsgemeinschaft within the SPP 1133 programme is greatfully acknowledged.
\section*{References}
\bibliographystyle{unsrt}

\end{document}